\documentclass[twocolumn,english,aps,prb,floatfix,amssymb,linenumbers]{revtex4}
\usepackage{amsmath}
\usepackage{dsfont}
\usepackage{color}
\usepackage{amssymb}
\usepackage{graphicx}
\usepackage{braket}
\usepackage{units}
\ifx\pdftexversion\undefined
\usepackage[dvips]{hyperref}
\else
\usepackage{hyperref}
\fi
\hypersetup{
  colorlinks = false
}

\newcommand{\ssm}{\scriptscriptstyle\rm}

\renewcommand{\theta}{\vartheta}

\begin{document}
\title{Observation of phononic helical edge states in a mechanical `topological insulator'}
\date{\today}
\author{Roman~S\"usstrunk}
\affiliation{Institute for Theoretical Physics, ETH Zurich, 8093 Z\"urich, Switzerland}
\author{Sebastian~D.~Huber}
\affiliation{Institute for Theoretical Physics, ETH Zurich, 8093 Z\"urich, Switzerland}

\begin{abstract}
A topological insulator is characterized by a dichotomy between the interior and the edge of a finite system: While the bulk has a non-zero energy gap, the edges are forced to sustain excitations traversing these gaps. Originally proposed for electrons goverened by quantum mechanics, it has remained an important open question if the same physics can be observed for systems obeying Newton's equations of motion. Here, we report on measurements that characterize the collective behavior of mechanical oscillators exhibiting the phenomenology of the quantum spin hall effect. The phononic edge modes are shown to be helical and we demonstrate their topological protection via the stability against imperfections. Our results open the door to the design of topological acoustic meta-materials that can capitalize on the stability of the surfaces phonons as reliable wave guides.
\end{abstract}

\maketitle

The experimental hallmarks of the quantum spin Hall effect (QSHE) in semi-conductor quantum wells\cite{Haldane88,Kane05,Bernevig06,Konig07,Moore10} are two counter-propagating edge modes that differ by their spin degree of freedom. As long as time reversal symmetry is preserved, these two modes are independent and do not scatter into each other. Much of the interest in condensed-matter research involving topological states is driven by the use of these protected edge modes for technological applications such as spintronics,\cite{Qi11,Hasan10} magnetic devices\cite{Mellnik14} or quantum information processing.\cite{Mong14} The transfer of the phenomenology of the QSHE from the quantum mechanical realm to classical mechanical systems is therefore both of fundamental interest and a gateway towards new design principles in mechanical meta-materials. 

Several key problems in the engineering of acoustic materials can potentially be addressed by capitalizing on the physics of the QSHE. The edge channels are robust counter-parts to the well known whispering gallery modes.\cite{Rayleigh89,Rayleigh10} Any application that requires energy to be confined to the surfaces of some device, for example vibration insulators, can potentially make use of such edge states. In contrast to the whispering gallery modes, which are extremely sensitive to the shape of the surface,\cite{Armani03} the topological edge modes are stable under a broad range of perturbations. Moreover, due to the stability of these modes, scattering-free phonon wave-guides of almost arbitrary shape can be realized. This in turn enables the engineering of robust acoustic delay lines\cite{Hafezi11,Hafezi13,Rechtsman13} useful for acoustic lensing.\cite{Spadoni10}

\begin{figure}[b!]
\includegraphics{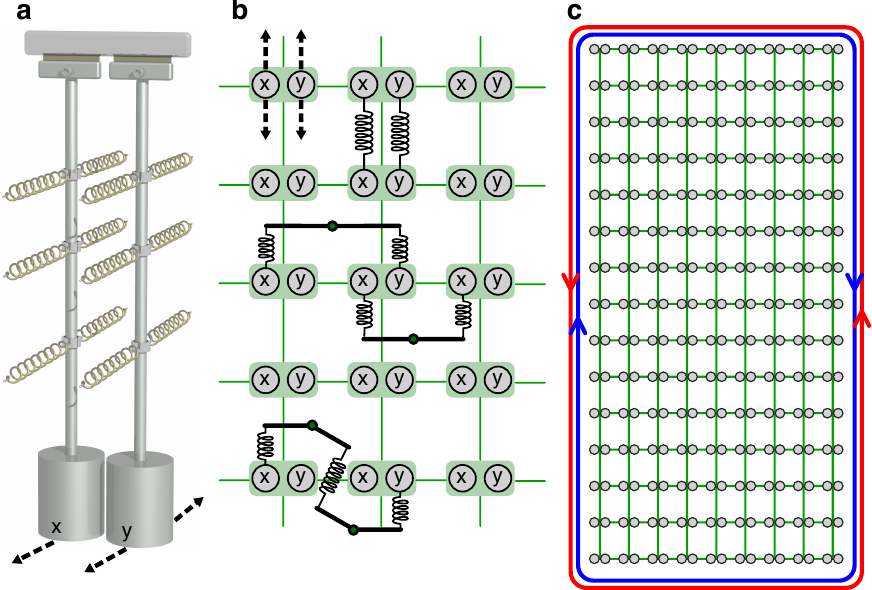}
\caption{
{\bf Setup.} {\bf a,} Illustration of two pendula making up one effective site of our lattice model. Six springs are attached to each pendulum for the coupling to its nearest neighbors.  {\bf b,} Schematic (top) view of the different types of couplings. Each green rectangle depicts a logical site hosting an ``$x$'' and a ``$y$'' pendulum swinging in the direction of the arrows. The longitudinal couplings are simple springs connecting the pendula. Transverse couplings are achieved via one or two lever arms, depending on the required sign of the coupling (see App.~\ref{app:dm}). {\bf c,} Our $9\times 15$ lattice made of 270 pendula. The two helical edge states known from the quantum spin Hall effect are indicated in blue and red.
}
\label{fig:setup}
\end{figure}

How a mechanical system described by Newton's equations can reproduce the phenomenology of a quantum mechanical model such as the QSHE has, however, remained an open question.\cite{Prodan09,Berg11,Kane13} Moreover, with a view to potential future applications, it is essential to demonstrate that the physics of the non-interacting and damping-free QSHE can be observed in a real mechanical system, which necessarily suffers from losses, disorder and non-linearities inherent to coupled mechanical oscillators.

\begin{figure*}[t!]
\begin{center}
\includegraphics{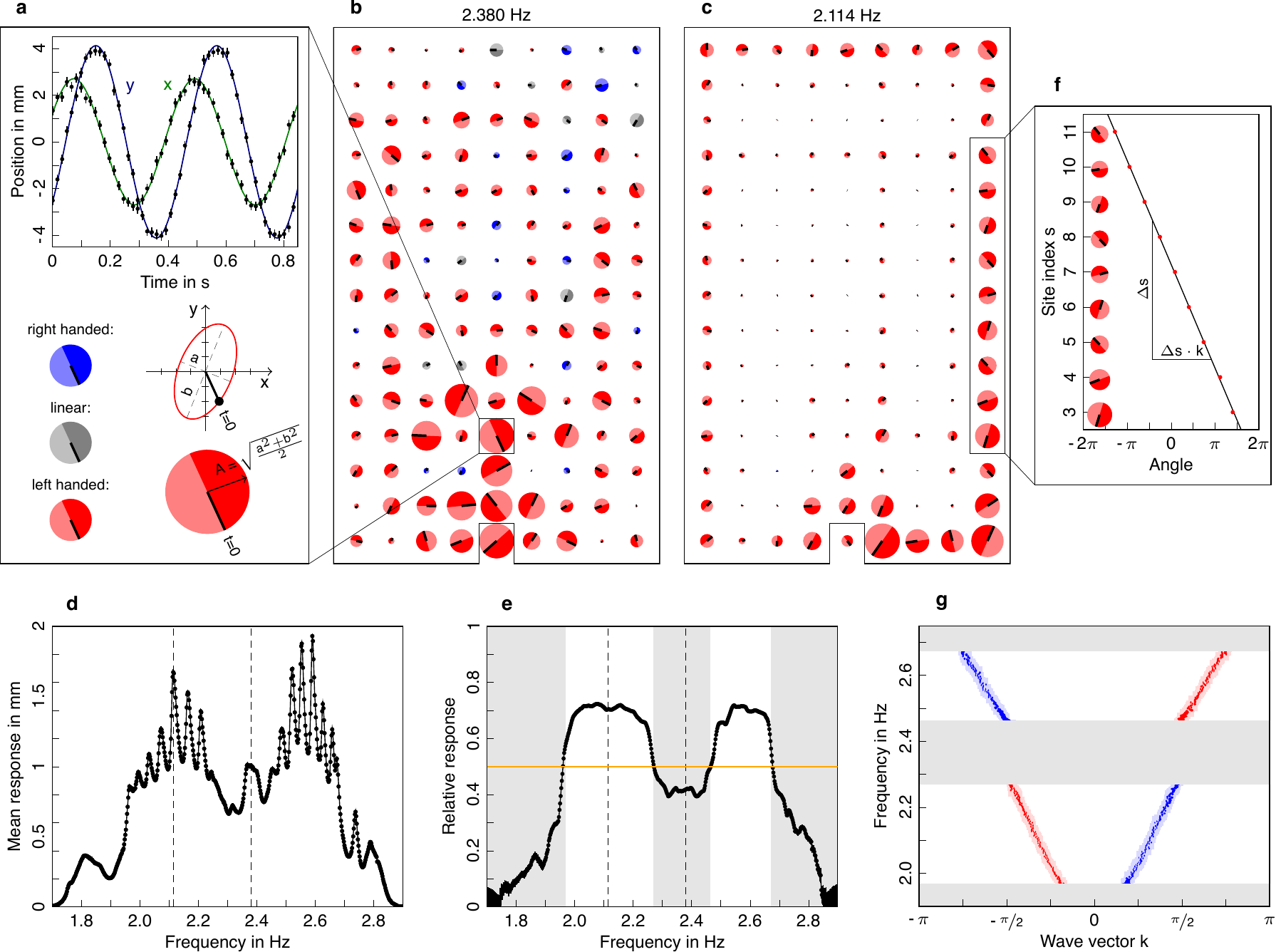}
\end{center}
\caption{
{\bf Helical edge mode dispersion.} {\bf a,} Time traces of two pendula. These can be interpreted as a two-dimensional trajectory, as shown. Steady states are displayed by colored disks representing their polarization. The radius of the circle corresponds to the mean deflection $A$ and the black line indicates the position of the pendula at a given fixed time. {\bf b \& c,} Measured steady states at two different frequencies ($2.114$ and $2.380\, {\rm Hz}$). The circles are normalized to the strongest deflection. The excluded site at the centre of the bottom row is excited with left-circular polarization. {\bf d,} Mean response of the system (average $A$) as a function of the excitation frequency showing an overall bandwidth of $\sim 1.2\, {\rm Hz}$. Error bars (smaller than the symbol size) are explained App.~\ref{app:data}. {\bf e,} The mean response of the edge relative to the bulk. In the frequency ranges shaded in grey, the bulk response dominates. The frequencies marked with dashed lines correspond to panels {\bf b \& c}. {\bf f,} For edge-dominated modes, the evolution of the angle indicated by the black line defines a wave vector $k$. {\bf g,} The frequency of the edge modes as a function of the wave vector. The bulk bands are shaded in gray. The color labels the polarization as before, establishing the helical nature of the edge excitations. The shape of the dark dots indicates their error (App.~\ref{app:data}) while the red (blue) shaded region marks a $0.04\,{\rm Hz}$ band corresponding to the loss-induced broadening of the eigen-frequencies.
}
\label{fig:spectra}
\end{figure*}

A quantum-mechanical lattice problem is described by a Schr\"odinger equation of the form
\par\nobreak\noindent
\begin{equation}
i\hbar \dot \psi_{i}^{\alpha} = \mathcal H_{ij}^{\alpha\beta} \psi_{j}^{\beta},
\end{equation}
\noindent
where $\psi_{i}^{\alpha}$ are the wave-function amplitudes for an electron with spin $\alpha$ on lattice site $i$ and $\mathcal H_{ij}^{\alpha\beta}$ is the Hamiltonian matrix describing the QSHE.\cite{Bernevig06a, Bernevig13} On the other hand, the dynamics of a collection of classical harmonic oscillators is described by Newton's equation of motion
\par\nobreak\noindent
\begin{equation}
\label{eqn:newton}
\ddot x_{i} =-\mathcal D_{ij} x_{j},
\end{equation}
\noindent
where $x_{i}$ are the coordinates of $N$ pendula ($i=1,\dots,N$) and $\mathcal D_{ij}$ is the dynamical matrix containing the couplings between them.
For either system, the existence and properties of the edge modes are features of the eigenstates of $\mathcal H$ or $\mathcal D$ alone and depend neither on the interpretation of $\psi_{i}^{\alpha}$ versus $x_{i}$, nor on the nature of the dynamics,  i.e., $i\partial_{t}$ versus $\partial_{t}^{2}$. Hence, we aim to design a dynamical matrix $\mathcal D$ incorporating the properties of the QSHE.

We start from two independent copies of the Hofstadter model\cite{Hofstadter76} on the square lattice with flux $\Phi=\pm \frac{2\pi}{3}$ per plaquette
\par\nobreak\noindent
\begin{equation}
\label{eqn:hofstadter}
\mathcal H = \!\!f\!\!\!\sum_{r,s,\alpha=\pm}\!\! \!\!|r,s,\alpha\rangle\langle r,s\pm1,\alpha| + |r,s,\alpha\rangle\langle r\pm 1,s,\alpha|e^{\pm i\alpha\phi_{s}}.
\end{equation}
\noindent
Here, $(r,s)$ denotes the location on the lattice of size $L_{r}\times L_{s}$, $\phi_{s}=|\Phi|s$ and $f$ represents the hopping amplitude. The two copies are labelled by a pseudo-spin index $\alpha=\pm$. $\mathcal H$ is symmetric under time reversal $\mathcal T$ and has three doubly degenerate bands, which are separated by non-zero gaps with a non-trivial topological $\mathds Z_{2}$ index, cf. App~\ref{app:band} \& \ref{app:chern}. Therefore, in a finite system we expect one chiral edge state per pseudo-spin in both gaps.

To take the step from the Hamiltonian $\mathcal H$ to the dynamical matrix $\mathcal D$, we need to ensure two properties. First, the eigenvalues of $\mathcal D$ are the squares of the eigenfrequencies of the dynamical system, cf. Eq.~(\ref{eqn:newton}). Therefore, $\mathcal D$ has to be positive definite, which can always be achieved by the addition of  a diagonal term $\propto \mu$. Second, $\mathcal D$ has to be strictly non-imaginary, as it encodes ideal springs. Owing to the time-reversal symmetry, this can be attained by a unitary transformation combining the local Kramers pairs $\alpha=\pm$
\par\nobreak\noindent
\begin{equation*}
\begin{pmatrix}
x_{r,s}^{\phantom{+}} \\
y_{r,s}^{\phantom{-}}
\end{pmatrix}=\underbrace{
\frac{1}{\sqrt{2}}
\begin{pmatrix}
1 & -i \\
1 & \phantom{-}i
\end{pmatrix}}_{=u}
\begin{pmatrix}
\psi_{r,s}^{+}\\
\psi_{r,s}^{-}
\end{pmatrix},
\qquad
U=u\otimes \mathds 1_{\rm lattice}.
\end{equation*}
\noindent
Each of the new coordinates $x_{r,s}$ and $y_{r,s}$ describes a one-dimensional oscillator at lattice site $(r,s)$. In the following, we interpret $(x_{r,s},y_{r,s})$ as the coordinates of a two-dimensional pendulum. It is then evident from the structure of $u$ that the eigenmodes of the system characterized by $\alpha=\pm$ correspond to left and right circularly polarized motions. Hence, these two polarizations replace the notion of pseudo-spin up and down of the quantum mechanical problem.
\begin{figure}[t]
\includegraphics{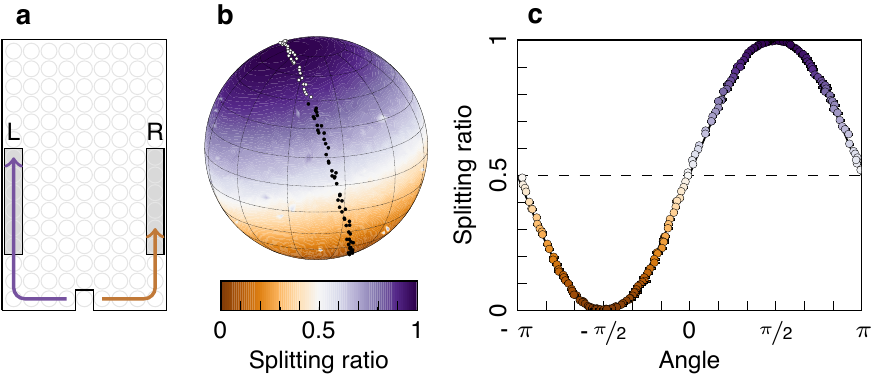}
\caption{
{\bf Beam splitter.} {\bf a,} Geometry of the beam splitter. An arbitrary polarization is injected at the indicated site at $2.123\,{\rm Hz}$. The relative weights $A$ on the two shaded boxes (R/L) define a splitting ratio. {\bf b,} Splitting ratio on the Bloch sphere. The north (south) pole of the sphere corresponds to left (right) circular polarizations of the drive and the color indicates the splitting ratio. The colormap is an interpolation based on 7000 data points.  The maximal splitting is not reached at the poles, indicating the presence of disorder. The dots mark measurements along a great circle through the points of maximal splitting. {\bf c,} Splitting ratio along the great circle. The black line marks a cosine expected for an optimal beam-splitter. The maximally reached splitting is $99.80(4)\%$.
}
\label{fig:splitter}
\end{figure}

We implement the dynamical matrix $\mathcal D=U^{\dag}\mathcal HU$ using pendula as depicted in Fig.~\ref{fig:setup}. $\mathcal D$ inherits its structure from the Landau gauge in Eq.~(\ref{eqn:hofstadter}): In $s$-direction, there is no cross-coupling between $x_{r,s}$ and $y_{r,s\pm 1}$ (no Peierls phase in $\mathcal H$). Along the $r$-direction, $\mathcal D$ also mixes the $x$- and $y$-degrees of freedom with a strength that depends on the $r$-coordinate (see App.~\ref{app:dm}). The total system consists of 270 pendula in a lattice of $L_{r}\times L_{s}=9 \times 15$ sites. The bare eigenfrequencies of the pendula due to gravity is $\omega_0/2\pi \approx 0.75\,{\rm Hz}$, which changes with the restoring forces of the spring couplings to $\omega/2\pi \approx 2.34\,{\rm Hz}$ (see App.~\ref{app:dm}).  

We harmonically excite one lattice site with a well-defined polarization by forcing the position of the two pendula. By tracking the position of all pendula one can extract the two-dimensional traces $[x_{r,s}(t),y_{r,s}(t)]$. From these we obtain the mean deflection $A_{r,s}$ as well as the polarization (left handed, linear or right handed) related to the lag between the $x$ and $y$ pendulum (see App.~\ref{app:data}), cf. Fig.~\ref{fig:spectra}(a). 

To establish the existence of the edge states, we scan the frequency at which one edge site is excited. For every frequency we wait until the system reaches a steady state before we extract the total response $\chi=\sum_{r,s}A_{r,s}/N$, where $N$ is the number of sites. Figure~\ref{fig:spectra}(d) shows that the system responds appreciably between $1.7\, {\rm Hz}$ and $2.9\, {\rm Hz}$, as expected (see App.~\ref{app:dm}). Moreover, there are two regions with sequences of pronounced peaks. The width of these peaks is $\Gamma \sim 0.04\,{\rm Hz}$, indicating the damping of the oscillators. In order to make the connection to edge modes we separate the system into the outermost line of lattice sites $\chi_{\rm e}=\sum_{\rm edge}A_{r,s}/N_{\ssm edge}$ and the rest $\chi_{\rm b}=\sum_{\rm bulk}A_{r,s}/N_{\ssm bulk}$. The relative weight $\chi_{e}/(\chi_{b}+\chi_{e})$ is shown in Fig.~\ref{fig:spectra}(e). There are three  bands where the response lies mainly in the bulk which are separated by two frequency regions where the response is dominated by the edge. To illustrate this further, we show in Fig.~\ref{fig:spectra}(b)~\&~(c) the recorded mode structure at a bulk and an edge frequency, respectively.

Given the analogy to the QSHE effect, we expect the edge states to be helical. The edge spectrum $\omega(k)$, where $k$ is the wave vector along the edge, can be extracted from the steady states shown in Fig.~\ref{fig:spectra}(c): Beyond $A_{r,s}$ and the polarization, we determine the position of each two-dimensional oscillator at a given time $[x_{r,s}(t_{0}),y_{r,s}(t_{0})]$. The angle of this vector with respect to the positive $x$-direction defines a local phase $\phi_{r,s}$, cf. Fig.~\ref{fig:spectra}(a). The change
\par\nobreak\noindent
\begin{equation*}
k=\phi_{r,s+1}-\phi_{r,s}
\end{equation*}
\noindent 
defines the wave vector in units of the inverse lattice constant. We show the resulting edge dispersions $\omega(k)$ in Fig.~\ref{fig:spectra}(g). For each polarization, there is a chiral (that is, uni-directional) mode per gap, as expected.
\begin{figure}[t!]
\begin{center}
\includegraphics{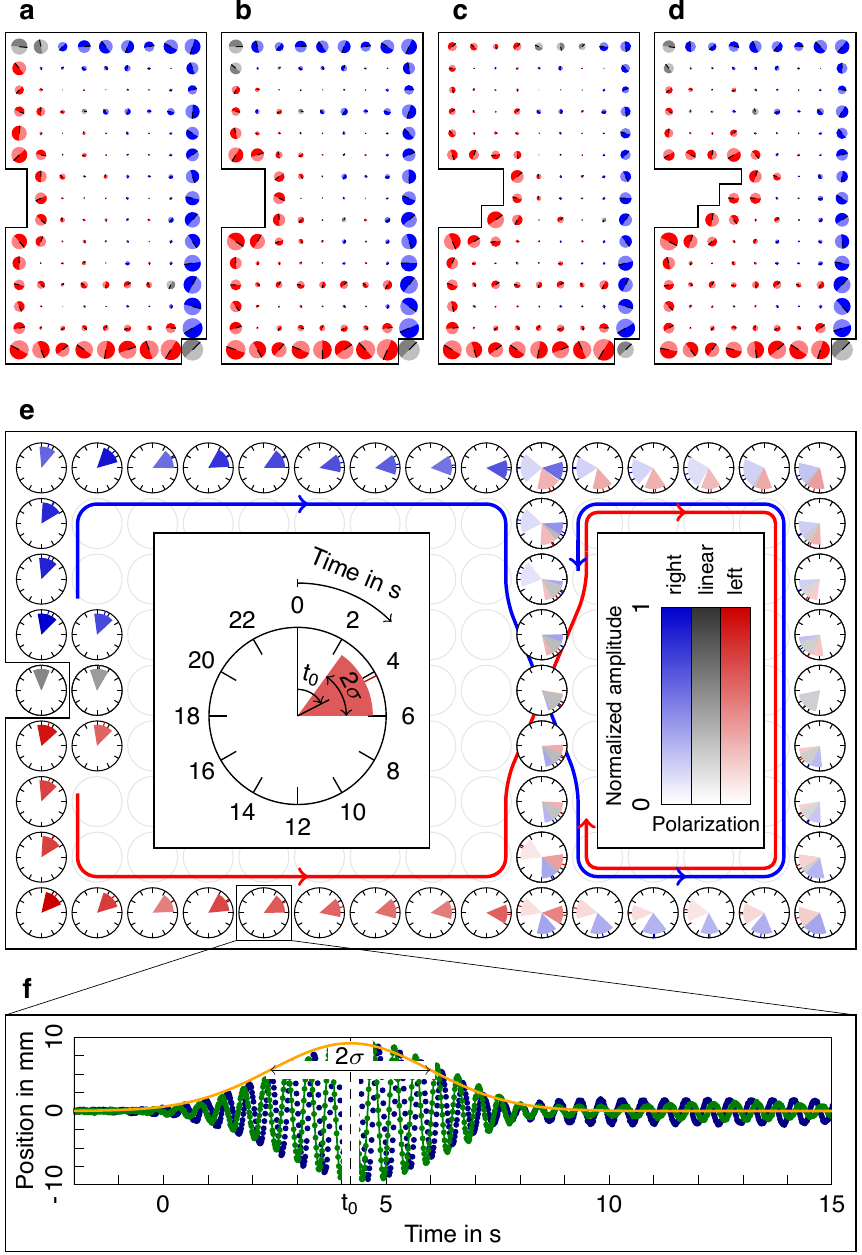}
\end{center}
\caption{
{\bf Topological protection.} {\bf a-d,} Steady states with a sequence of removed sites illustrating the stability of the edge states against boundary roughness. The bottom right site is driven with linear polarization at a frequency in the upper gap. {\bf e,} Wave packets launched at the marked site with linear polarization at a frequency in the lower gap. Each clock represents an edge site on which the blue (red) wedge is centered at the time $t_{0}$ when the wave packet traverses the site. The angle of the wedge indicates the width $\sigma$ of the wave packet as illustrated in panel f. The color represents its polarization. From these clocks one can read off the propagation of the wave-packets throughout the system. {\bf f,} Analysis of the wave-packet at a given site from which the passing time $t_{0}$ and the width of the wave-packet $\sigma$ are extracted.
}
\label{fig:wp}
\end{figure}

The edge dispersion indicates that the classical system faithfully implements the QSHE. However, the edge state of the QSHE are protected by the symmetry $\mathcal T$. In the mechanical case, this symmetry is not related to time reversal. Instead $\mathcal D$ is only symmetric under $\mathcal T$ if the ratios of the local couplings are correctly implemented. Hence, disorder in the spring constants induces both $\mathcal T$-breaking terms that can remove the edge modes as well as benign disorder effects which only remove left- and right circular polarization as a global integral of motion. 

For a clean system, a linearly polarized excitation on a boundary site is split into two counter-propagating circularly polarized modes. Therefore the edge states can be used as a polarizing beam splitter. $\mathcal T$-symmetry breaking diminishes the efficiency of the beam splitter by coupling the two polarizations. $\mathcal T$-symmetric disorder, on the other hand, only leads to a selection of different polarizations in the splitting process. 

We measured the ratio of excitation $\sum_L A_{r,s}/\sum_R A_{r,s}$ between the left and the right long edge after exciting in the middle of a short edge (Fig.~\ref{fig:splitter}(a)). We scanned all possible polarizations by changing the relative amplitude and time-lag between the two local pendula. In analogy to the quantum mechanical case, we plot the splitting ratio as a colormap on a Bloch sphere, where the north pole (south pole) corresponds to right (left) circular polarization of the drive, respectively. The maximal imbalance between the left and right edge are not reached on the north (south) pole, but on two approximately antipodal  points rotated by $\sim\!15^{\circ}$ away from the poles (see Fig.~\ref{fig:splitter}(b)). However, as illustrated in Fig.~\ref{fig:splitter}(c), we reach a splitting fidelity of $99.80(4)\%$ at the optimal points. From this we conclude that on the length-scale of our system, only  disorder symmetric under time reversal is relevant (see App.~\ref{app:tr}).

Surface states in mechanical systems with concave boundaries are known as whispering gallery modes.\cite{Rayleigh89,Rayleigh10} To demonstrate that our edge states are not mere whispering gallery modes and to highlight the robustness of the phononic edge states described here, we remove a sequence of sites from the dynamical problem, effectively creating a convex boundary. In Fig.~\ref{fig:wp}(a)--(d), we show the resulting mode structure for a frequency in the lower gap. The result demonstrates that the exact shape of the boundary has no influence on the stability of the edge states, in contrast to whispering-gallery modes, which are exquisitely sensitive to imperfections. 

To further strengthen the point that the edge states are topological rather than imposed by the finite-size geometry, we create a domain wall between two different topological sectors. We invert the effective flux seen by the two polarizations on six rows of the system (see App.~\ref{app:dm}). At the boundary between the two sectors, the spin Chern numbers change their values, which requires the presence of in-gap modes. We illustrate these topological in-gap modes along the sector boundaries by exciting a linearly polarized wave-packet on the short edge of the larger sector. Thereby we also demonstrate the unique ability, owing to the long timescales in our experiments, to investigate dynamical rather than steady-state properties. As shown in Fig.~\ref{fig:wp}(e), the wave packet splits into two circularly polarized packets, each traveling along one edge. At the boundary to the second sector, they are deflected into the interior of the lattice (where they interfere to a linearly polarized wave-packet again), before they each travel independently along the physical edge of the second sector in reversed directions.

In this work, we have realized a mechanical `topological insulator' displaying the phenomenology of the QSHE. Our results show that the topologically protected edge states are observable in a  mechanical system despite the presence of disorder and decay processes. For potential technological applications, the ability to define `edge' states along domain-walls between different topological sectors is particularly promising. By stacking different domains, where the length of the domain walls can be controlled by the precise shape of the interface, stable acoustic delay lines can be engineered.\cite{Spadoni10} Moreover, the work presented here establishes the bridge between the quantum mechanical phenomena of topological insulators and mechanical systems. The outlook to implement the acoustic counter-part of three-dimensional topological insulators is peculiarly interesting.

 We would like to thank T. Donner, T. Esslinger, P. Maletinsky, E.P.L. van Nieuwenburg, M. Tovmasyan, and O. Zilberberg for discussions. Special thanks go to C. Daraio for pointing out the technological relevance of our work. We acknowledge financial support from the Swiss National Foundation.

\bibliographystyle{phd-url}
\bibliography{ref}

\appendix

\section{Dynamical matrix in real space}
\label{app:dm}
Starting from Eq. (\ref{eqn:hofstadter}) we apply the unitary transformation $U$ to obtain $D = U^{\dagger} \mathcal H U$. The resulting equations of motion read explicitly
\par\nobreak\noindent
\begin{align}
\label{eqn:a1}
\ddot x_{r,3s} & = - (\omega_{0}^{2}+Af)x_{r,3s}+f(x_{r\pm1,3s}+x_{3r,3s\pm1}),\\
\ddot y_{r,3s} & = - (\omega_{0}^{2}+Af)y_{r,3s}+f(y_{r\pm1,3s}+y_{3r,3s\pm1}),\\
\ddot x_{r,3s+1} & = -(\omega_{0}^{2}+Af)x_{r,3s+1}+fx_{r,3s+1\pm 1}\\
\nonumber
&-\frac{f}{2}(x_{r\pm1,3s+1}) + \frac{\sqrt{3}f}{2}(y_{r+1,3s+1}-y_{r-1,3s+1}),\\
\ddot y_{r,3s+1} & = - (\omega_{0}^{2}+Af)y_{r,3s+1}+fy_{r,3s+1\pm 1}\\
\nonumber
&-\frac{f}{2}(y_{r\pm1,3s+1}) + \frac{\sqrt{3}f}{2}(-x_{r+1,3s+1}+x_{r-1,3s+1}),\\
\ddot x_{r,3s+2} & = - (\omega_{0}^{2}+Af)x_{r,3s+2}+fx_{r,3s+2\pm 1}\\
\nonumber
&-\frac{f}{2}(x_{r\pm1,3s+2}) + \frac{\sqrt{3}f}{2}(-y_{r+1,3s+2}+y_{r-1,3s+2}),\\
\ddot y_{r,3s+2} & = - (\omega_{0}^{2}+Af)y_{r,3s+2}+fy_{r,3s+2\pm 1}
\label{eqn:a2}
\\
\nonumber
&-\frac{f}{2}(y_{r\pm1,3s+2}) + \frac{\sqrt{3}f}{2}(x_{r+1,3s+2}-x_{r-1,3s+2}).
\end{align}
\noindent
Here, the term proportional to $A=3\!+\!\sqrt{3}$ encodes the restoring acceleration arising from the springs described by $f=M/J$, where $J$ is the moment of inertia and $M$ the torque constant of the springs. In our implementation with pendula of $500\,{\rm mm}$ length and a mass of approximately $500\,{\rm g}$ gravity gives rise to a bare frequency of $\omega_{0}/2\pi\approx 0.75\, {\rm Hz}$. The coupling constants are chosen to give $\sqrt{f}/2\pi\approx 1.02\, {\rm Hz}$. The above equations of motions are implemented as outlined in Fig.~\ref{fig:setup}. For couplings with a positive sign in Eqns.~(\ref{eqn:a1}--\ref{eqn:a2}) in the $s$-direction simple springs between nearest neighbors are needed. For positive couplings in $r$-direction, two-lever arms connecting the pendula with springs achieve the same effect: A deflection of one pendulum induces a deflection of the second in the same direction. For negative couplings, this direction can be inverted via the use of only one lever arm as shown in Fig.~\ref{fig:setup}. We chose all lever arms and springs light enough such that their inertia is negligible. The lattice spacing is 135~mm in $r$-direction and 120~mm in $s$-direction. Note that the effect of interchanging the flux $\Phi\rightarrow -\Phi$ in $\mathcal H$ amounts to an interchange of the sign in the cross-coupling between $x$ and $y$ pendula. Therefore, to create two different topological sectors one inverts these couplings on parts of the system.

\section{Band structure} 
\label{app:band}

$\mathcal H$ is symmetric under time reversal $\mathcal T=i\sigma_{y}\mathcal K$, where $\sigma_{y}$ is the Pauli matrix acting on the $\alpha$-index and $\mathcal K$ denotes complex conjugation. $\mathcal T^{2}=-1$ which puts $\mathcal H$ in symmetry class AII.\cite{Schnyder08} In two dimensions this class contains the QSHE and is characterized by a $\mathds Z_{2}$ topological index.\cite{Kitaev09}

Both pseudo-spin sectors of $\mathcal H$ can be diagonalized and yield each three bands with dispersion
\par\nobreak\noindent
\begin{equation*}
\epsilon_{\gamma}(k_{r},k_{s}) = \mu + f\frac{2e^{ 2\pi i \gamma/3}+e^{- 2\pi i \gamma/3}\beta(k_{r},k_{s})^{2/3}}{\beta(k_{r},k_{s})^{1/3}},
\end{equation*}
\noindent
where 
\par\nobreak\noindent
\begin{equation*}
\beta(k_{r},k_{s}) = -\cos(q_{r})-\cos(q_{s})+\sqrt{[\cos(q_{r})+\cos(q_{s})]^{2}-8}.
\end{equation*}
\noindent
Here $\gamma=0,1,2$ and $q_{r(s)}$ are the momenta in $r$ ($s$) direction, respectively. For our parameters the eigenfrequencies are given by $\omega_{\gamma}(q_{k},q_{s})=\sqrt{\epsilon_{\gamma}(q_{r},q_{s})}$ which are bounded by ${\rm min}[\omega_{\gamma}(q_{r},q_{s})]\approx1.64\,{\rm Hz}$ and ${\rm max}[\omega_{\gamma}(q_{r},q_{s})]\approx 2.91\,{\rm Hz}$. 

\section{Spin Chern numbers} 
\label{app:chern}

In the pure case $\mathcal H$ describes two copies of the Hofstadter problem of a particle hopping on a square lattice in the presence of a flux $\Phi=\frac{2\pi}{3}$ per plaquette. Owing to the absence of any coupling between the two copies, a topological index can be defined via the spin Chern numbers ${\bf C}=(C_{+},C_{-})$ evaluated for each sector separately.\cite{Sheng06} The standard\cite{Kane05} $\mathds Z_{2}$ index $\nu$ is then given by
\par\nobreak\noindent
\begin{equation}
\nu = \frac{C_{+}-C_{-}}{2} \,{\rm mod}\, 2.
\end{equation}
\noindent
In the present case the two copies $\alpha=\pm$ have opposite flux and accordingly $C_{+}=-C_{-}$. The Chern numbers for each gap can be calculated from the Diophantine equation\cite{Thouless82} and yield 
$C_{+}=1$ for the lower and $C_{+}=-1$ for the upper gap yielding a non-trivial $\nu=1$. Any change in the Chern number is accompanied by a gap closing. Hence, the boundary between the different sectors of flux $\Phi=\pm \frac{2\pi}{3}$, where $C_{\pm}$ change by $\pm 2$ carries two co-propagating edge channels per ``spin''.

\section{Data analysis} 
\label{app:data}

Time traces of all 270 pendula are measured in parallel with two cameras at an acquisition rate of 60~Hz. Due to the finite resolution of the cameras, the resulting position has an intrinsic error of $0.2\,{\rm mm}$. Typical deflections are in the range of a few millimeter. This amounts to a relative measurement error between $10^{-1}$ and $10^{-2}$. Moreover, the relation of the distance measured on the camera and the angle of deflection are in linear relation to a good approximation; we use mm throughout the text. We fit the obtained traces to the function
\par\nobreak\noindent
\begin{multline}
\begin{pmatrix}
x_{r,s}(t)\\
y_{r,s}(t)
\end{pmatrix}=
A_{r,s}\bigg\{\sin\left[\frac{1}{2}\left(\varphi_{r,s}+\frac{\pi}{2}\right)\right]
\begin{pmatrix} 
\sin(\omega t) \\ \cos(\omega t)
\end{pmatrix}
\\+
\cos\left[\frac{1}{2}\left(\varphi_{r,s}+\frac{\pi}{2}\right)\right]
\begin{pmatrix} 
\sin(-\omega t+\theta_{r,s}) \\ \cos(-\omega t+\theta_{r,s})
\end{pmatrix}\bigg\}.
\end{multline}
\noindent
The resulting mean displacements $A_{r,s}$, the polarizations $\varphi_{r,s}$, and the angles $\theta_{r,s}$ have typically relative errors of maximally $10^{-3}$ owing to the large number of measurements. Positive (negative) values of $\varphi_{r,s}$ correspond to left and right circular or elliptic polarizations. For $\varphi_{r,s}=0$, the angle $\theta_{r,s}$ corresponds the orientation of the linear polarization. In the general case $\varphi_{r,s}$ and $\theta_{r,s}$ parameterize the polar and azimuthal angle of the Bloch sphere shown in Fig.~\ref{fig:splitter}.

\section{Localization length}
\label{app:tr}

In order to assess the effect of $\mathcal T$-symmetry breaking terms we calculated the localization length $\xi$. We used the Kwant\cite{Groth14} package to determine the longitudinal conductivity $\sigma(L)$ of $\mathcal H$ on a finite system of length $L$ under the influence of disorder. The springs used in the experiments have an uncertainty of $\sim 4\%$ in their stiffness described by $f$ in Eqns.~(\ref{eqn:a1}--\ref{eqn:a2}). We assume Eqns.~(\ref{eqn:a1}--\ref{eqn:a2}) to be disordered accordingly, transform back to the quantum mechanical problem $\mathcal H$ and extract $\xi$ from the scaling behavior of $\sigma(L)$ with system size. The resulting localization length depends on energy. For states in the gap we find the localization length of the edge states to be  $\xi \approx 1000$ lattice constants. For our system with a maximal distance along the edge of 46 sites, we therefore expect $\mathcal T$-symmetry breaking to play no essential role for the edge channels. 

\end{document}